\documentclass[12pt,english,floatfix,superscriptaddress,aps,prd,preprint]{revtex4}
\usepackage[utf8]{inputenc}
\usepackage{float}
\usepackage{array}
\usepackage{lipsum}
\usepackage{graphicx}
\usepackage{amsmath,amsthm,amsfonts,amssymb}
\usepackage{graphicx}
\usepackage[english]{babel}
\usepackage{color}
\usepackage{esint}
\usepackage[dvips]{epsfig}
\usepackage[dvips]{graphicx}
\usepackage{float}
\usepackage{units}
\usepackage{textcomp}
\usepackage{mathrsfs}
\usepackage{amsmath}
\usepackage[makeroom]{cancel}
\usepackage{amssymb}
\usepackage{amsbsy}
\usepackage{amsfonts}

\usepackage{hyperref}
\usepackage{slashed}

\newcommand{\ie}{\begin{equation}}
\newcommand{\fe}{\end{equation}}
\newcommand{\se}{\begin{eqnarray}}
\newcommand{\ff}{\end{eqnarray}}

\begin{document}

\title{Antisymmetric tensor propagator with spontaneous Lorentz violation}

\author{R. V. Maluf}
\email{r.v.maluf@fisica.ufc.br}
\affiliation{Universidade Federal do Cear\'a (UFC), Departamento de F\'isica,\\ Campus do Pici, Fortaleza - CE, C.P. 6030, 60455-760 - Brazil.}


\author{A. A. Ara\'{u}jo Filho}
\email{dilto@fisica.ufc.br}
\affiliation{Universidade Federal do Cear\'a (UFC), Departamento de F\'isica,\\ Campus do Pici, 
Fortaleza - CE, C.P. 6030, 60455-760 - Brazil.}


\author{W. T. Cruz}
\email{wilamicruz@gmail.com}
\affiliation{Instituto Federal de Educa\c{c}\~ao, Ci\^encia e Tecnologia do Cear\'a (IFCE),
Campus Juazeiro do Norte, 63040-000 Juazeiro do Norte - CE - Brazil }


\author{C. A. S. Almeida}
\email{carlos@fisica.ufc.br}
\affiliation{Universidade Federal do Cear\'a (UFC), Departamento de F\'isica,\\ Campus do Pici, Fortaleza - CE, C.P. 6030, 60455-760 - Brazil.}

\date{\today}

\begin{abstract}
In this work, we study the spontaneous Lorentz symmetry breaking due to an antisymmetric 2-tensor field in Minkowski spacetime. For a smooth quadratic potential, the spectrum of the theory exhibits massless and massive excitations. We show that the equations of motion for the free field obey some constraints which lead to the massive mode be non-propagating at leading order. Besides, there exists a massless mode in the theory which can be identified with the usual Kalb-Ramond field, carrying only one on-shell degree of freedom. The same conclusion holds when one analyses the pole structure of its Feynman propagator. A new complete set of spin-type operators is found, which was the requirement to evaluate the propagator of the Kalb-Ramond field modified by the presence of a nonzero vacuum expectation value responsible for the Lorentz violation.
\end{abstract}

\maketitle

\section{Introduction}

In the last years, the possibility of CPT and Lorentz symmetry violations has been intensively investigated in the context of the standard model extension (SME) \cite{SME1}. The SME is conceived as an effective field theory that accounts for the Lorentz violating effects and preserves the gauge structure of the elementary particles and  the fundamental interactions described by the standard model (SM) \cite{Pott1}. In the framework of the SME, the violation of Lorentz symmetry can be implemented by two distinct ways in flat spacetime, either explicitly or dynamically. 
The explicit Lorentz invariance violation (LIV) can be accomplished by adding Lorentz-violating coefficients directly in the Lagrangian density of the SM. However, it has been shown that the explicit breaking is incompatible with geometric identities such as the Bianchi identity in Riemann geometry, and therefore a convenient mechanism for addressing the LIV would be through spontaneous breaking within the gravitational sector \cite{KosteleckyG1}.

Field theories which involve spontaneous Lorentz symmetry breaking may be built up from $p$-forms, including vectors and antisymmetric $p$-tensors, which acquire nonzero vacuum expectation values triggered by a potential that plays some interesting features to the spectrum of the theory \cite{BluhmFung}. This potential can take different forms, such as the smooth quadratic \cite{Bluhm2005,BluhmFung,MalufGravity1,MalufGravity2,TMarizRadiative,MalufRadiative}, Lagrange-multiplier \cite{BluhmFung} and nonpolynomial \cite{altschul}. In addition, the theory may contain field excitations around the vacuum solution which can be classified as Nambu–Goldstone (NG), massive, Lagrange-multiplier, and spectator modes  \cite{BluhmFung}. Amongst the possible choices for the models with a gauge-invariant kinetic term, special attention is given to the case involving antisymmetric $p$-tensor fields \cite{AntiSy-Ogievetsky}. The main example of this type of theories involves a gauge invariant kinetic term for an antisymmetric 2-tensor field, commonly called of the Kalb-Ramond field \cite{KalbRamond}. When nonminimal curvature couplings are present, the Kalb-Ramond field can be used to describe the dynamics of all LIV coefficients in the gravitational SME sector, usually denoted by $u$, $s^{\mu\nu}$, and $t^{\mu\nu\kappa\lambda}$. The physical content and the phenomenological implications of this theory in Minkowski and Riemann spacetimes were first analyzed in Ref. \cite{AntiSy-Altschul}.

In this work, we revisit the field theory that describes spontaneous Lorentz symmetry breaking due to a non-zero $vev$ for an antisymmetric 2-tensor $B_{\mu\nu}$ in Minkowski spacetime. We begin with the analysis of the free propagation modes for fluctuation field $\tilde{B}_{\mu\nu}$ at tree level. We show that the massive longitudinal excitation does not represent a physical propagating mode. Next, we calculate the exact free propagator for the Kalb-Ramond field modified by the presence of the LIV background tensor $b_{\mu\nu}$ applying the method of spin-projection operators \cite{Barnes,Sezgin,Boldo}. To accommodate the emerging Lorentz-violating terms, we extended the usual basis of projectors to find a closed algebra for a new set of spin-type operators. In agreement with the results previously found, no massive physical pole was generated.

This work is outlined as follows: Firstly, we introduce the model reviewing the main properties of the Kalb-Ramond field with spontaneous Lorentz violation and study its free propagation in the absence of matter.
Secondly, we calculate the modified Kalb-Ramond propagator for a smooth quadratic potential. Finally, we present comments about our results. In the present work, we adopt the metric signature as $(+1,-1,-1,-1)$.

\section{Theoretical model and the physical spectrum\label{themodel}}

Let us start by defining the Lagrangian density that describes the dynamics for an antisymmetric 2-tensor $B_{\mu\nu}$  in 4D Minkowski spacetime,
\ie
\mathcal{L} = \frac{1}{6} H_{\mu\nu\alpha}H^{\mu\nu\alpha} - V + B_{\mu\nu}J^{\mu\nu},\label{1}
\fe where
\ie H_{\mu\nu\alpha}=\partial_{\mu}B_{\nu\alpha} + \partial_{\alpha}B_{\mu\nu} + \partial_{\nu}B_{\alpha\mu},\label{2}\fe
is the field strength tensor associated with $B_{\mu\nu}$, $V$ is the potential that gives rise to the spontaneous Lorentz violation, and $J^{\mu\nu}$ is an antisymmetric conserved current due to the coupling to the matter \cite{AntiSy-Altschul}.  
The field strength $ H_{\mu\nu\alpha}$ satisfies the identity
\ie
\partial_{\kappa}H_{\lambda\mu\nu} - \partial_{\lambda}H_{\mu\nu\kappa} + \partial_{\mu}H_{\nu\kappa\lambda} - \partial_{\nu}H_{\kappa\lambda\mu} = 0,
\label{3}\fe and it is invariant under the gauge transformation of $B_{\mu\nu}$,
\ie
B_{\mu\nu}(x) \rightarrow B'_{\mu\nu}(x)=B_{\mu\nu}(x) + \partial_{\mu} \Lambda_{\nu}(x) - \partial_{\nu} \Lambda_{\mu}(x), \label{4}
\fe
where $\Lambda_{\mu}$ is an arbitrary vector field.  The field $\Lambda_{\mu}$ also exhibits an extra gauge invariance given by 
\ie
\Lambda_{\mu}(x) \rightarrow \Lambda'_{\mu}(x)= \Lambda_{\mu}(x) + \partial_{\mu} \Sigma(x),\label{5}
\fe with $\Sigma$ being an arbitrary scalar field. This latter transformation leaves Eq. \eqref{4} unchanged.

Before analyzing the effects of the spontaneous Lorentz symmetry breaking on the $B_{\mu\nu}$ field, it is instructive to review how this field behaves in the case when $V=0$, i.e., when the theory described in \eqref{1} has gauge symmetry. It is known that the Lagrangian density \eqref{1} describes a theory with only one physical degree of freedom, being equivalent to a real scalar field. One way to see this is using gauge symmetry to eliminate spurious components of $B_{\mu\nu}$ field by convenient choices of parameters associated with gauge transformations. 

For simplicity, let us assume $J^{\mu\nu}=0$, such that the equation of motion obtained from $\mathcal{L}$ are written as
\ie
\Box B_{\mu\nu}+\partial_{\mu}\partial^{\alpha}B_{\nu\alpha}+\partial_{\nu}\partial^{\alpha}B_{\alpha\mu}=0,\label{eqBox1}\fe and by simple inspection, we can verify that they are invariant under the gauge transformation \eqref{4}. Thus, we can not expect to obtain a unique solution for \eqref{eqBox1}, since we can always generate a new solution by performing a transformation in the form \eqref{4}. This ambiguity can be removed by choosing a particular gauge. A simple choice, analogous to the Lorentz gauge in electrodynamics, is 
\ie
\partial_{\mu}B^{\mu\nu}=0,\label{GC1}
\fe such that the equation of motion becomes
\ie
\Box B_{\mu\nu}=0.\label{EMotion1}
\fe 

Due to its antisymmetry, $B_{\mu\nu}$ has six independent components, but not all represent physical degrees of freedom, and some of them can be eliminated by the gauge condition \eqref{GC1}. However, this condition does not suffice to fix entirely the gauge choice and represents only three independent constraints that the components of the Kalb-Ramond field must satisfy. Indeed, we can still construct a solution $B'_{\mu\nu}=B_{\mu\nu}+\partial_{\mu}\Lambda_{\nu}-\partial_{\nu}\Lambda_{\mu}
 $, which preserves \eqref{GC1} and satisfying the equation of motion \eqref{EMotion1}, since $\Box\Lambda_{\mu}=0$ with $\partial_{\mu}\Lambda^{\mu}=0$. Such a solution is possible due to the residual gauge symmetry \eqref{5}, by setting the gauge parameter $\Sigma$ to satisfy the harmonic condition $\Box\Sigma=0$. 
 
Thus, the gauge parameter $\Lambda_{\mu}$, analogously to the usual gauge field $A_{\mu}$ in electrodynamics, has only two independent components which can be configured to reduce from three to one component of $B'_{\mu\nu}$, resulting simply in one single physical degree of freedom.

In general, the potential $V$ could incorporate dependence on $B_{\mu\nu}$, derivatives of $B_{\mu\nu}$, the metric $\eta_{\mu\nu}$, and the Levi-Civita tensor $\epsilon_{\mu\nu\alpha\beta}$. For the sake of simplicity, we will examine the content of the theory defined by \eqref{1} when $V$ takes one specific form. In what follows, our attention will be focused only on the kinetic and potential parts of the Lagrangian density \eqref{1}, such that the matter coupling, represented by $J_{\mu\nu}$, will be considered zero hereafter.

The simplest case for triggering the spontaneous Lorentz symmetry breaking is when the potential takes the form $V=V(X)$, with $X\equiv B_{\mu\nu}B^{\mu\nu} - b_{\mu\nu}b^{\mu\nu}$ such that the potential assume a minimum with a nonzero vacuum expectation value for $B_{\mu\nu}$,
\ie
\langle B_{\mu\nu} \rangle \equiv b_{\mu\nu}.\label{6}
\fe

More specifically, we choose a smooth quadratic potential in which the density Lagrangian \eqref{1} takes the form
\ie
\mathcal{L}_{B,V} = \frac{1}{6} H_{\mu\nu\lambda}H^{\mu\nu\lambda} - \frac{1}{2} \lambda ( B_{\mu\nu} B^{\mu\nu} - b^{2})^{2}, \label{LBV}
\fe where $\lambda$ is a dimensionless positive constant and $b^{2}\equiv b^{\mu\nu}b_{\mu\nu}$. 

For the theory described by $\mathcal{L}_{B,V}$, we are interested in studying the behaviour of $B_{\mu\nu}$ around the vacuum expected value $b_{\mu\nu}$. Thus, let us assume the decomposition
\ie
B_{\mu\nu} = b_{\mu\nu}+\tilde{B}_{\mu\nu},\label{vacuum}
\fe where $\tilde{B}_{\mu\nu}$ is the vacuum fluctuation, and $b_{\mu\nu}$ satisfies the requirement $\partial_{\mu}b^{\mu\nu}=0$. This assumption guarantees the translational invariance of the vacuum state and consequently the conservation of energy-momentum. Moreover, it is worth mentioning that $V(X)$ has five NG and one massive modes regarding field excitation  $\tilde{B}_{\mu\nu}$ \cite{AntiSy-Altschul}.

Thus, according to expansion \eqref{vacuum} we can rewrite $\mathcal{L}_{B,V}$ in the form
\begin{eqnarray}
\mathcal{L}_{B,V}&=&\frac{1}{6}\tilde{H}_{\mu\nu\alpha}\tilde{H}^{\mu\nu\alpha}
-2\lambda b_{\mu\nu}b_{\alpha\beta}\tilde{B}^{\mu\nu}\tilde{B}^{\alpha\beta}
\nonumber\\
&-&2\lambda b_{\alpha\beta} \tilde{B}^{\alpha\beta} \tilde{B}_{\mu\nu}\tilde{B}^{\mu\nu}-\frac{1}{2}\lambda(\tilde{B}_{\mu\nu}\tilde{B}^{\mu\nu})^{2},\label{Lexpan}
\end{eqnarray}where $\tilde{H}_{\mu\nu\alpha}$ is a field strength for $\tilde{B}_{\mu\nu}$. In the first line of the above expression, we notice that the presence of a mass term for $\tilde{B}_{\mu\nu}$ is described by the mass matrix $m_{\mu\nu,\alpha\beta}=4\lambda b_{\mu\nu}b_{\alpha\beta}$. The last line of \eqref{Lexpan} describes cubic and quartic self-interactions induced by the spontaneous breaking. 

In the present work, our main interest is studying the free propagation of the $\tilde{B}_{\mu\nu}$ field in the absence of matter. For this purpose, we focus our attention only on the bilinear terms of \eqref{Lexpan} which yield the equation of motion 
\ie
\partial_{\mu}\tilde{H}^{\mu\nu\alpha} + 4\lambda\tilde{B}_{\rho\sigma}b^{\rho\sigma} b^{\nu\alpha} = 0.  
\label{eqmotion3}\fe
The solutions of Eq. \eqref{eqmotion3} contain both massless NG and massive modes appeared mixed by the mass matrix $m_{\mu\nu,\alpha\beta}$. To separate these modes and reveal the physical content of the theory, we introduce transverse and longitudinal projectors concerning the orientation defined by $b_{\mu\nu}$:
\ie
P^{\parallel}_{\mu\nu , \alpha\beta} \equiv \frac{b_{\mu\nu}b_{\alpha\beta}}{b^{2}} \ \ \ \mbox{and} \ \ \ P^{\perp}_{\mu\nu , \alpha\beta} \equiv \mathcal{I}_{\mu\nu , \alpha\beta} - P^{\parallel}_{\mu\nu , \alpha\beta},
\label{projectors1}\fe 
where $\mathcal{I}_{\mu\nu , \alpha\beta}$ is the identity operator for rank-2 antisymmetric tensors, and it is defined as 
\ie
\mathcal{I}_{\mu\nu , \alpha\beta}=\frac{1}{2}(\eta_{\mu\alpha}\eta_{\nu\beta}-\eta_{\mu\beta}\eta_{\nu\alpha}).
\fe
Thus, the excitation $\tilde{B}_{\mu\nu}$ can be written in terms of the transverse and longitudinal components  in the following way: 
\begin{eqnarray}
\tilde{B}_{\mu\nu}	&=&	A_{\mu\nu}+\beta\hat{b}_{\mu\nu},\\
A_{\mu\nu} &	\equiv & 	P_{\mu\nu,\alpha\beta}^{\perp}\tilde{B}^{\alpha\beta}\ \ \ \mbox{(transverse mode),}\\
\beta\hat{b}_{\mu\nu} &\equiv &	P_{\mu\nu,\alpha\beta}^{\parallel}\tilde{B}^{\alpha\beta}\ \ \ \mbox{(longitudinal mode),}
\end{eqnarray} with $A_{\mu\nu}b^{\mu\nu}=0$, $\beta=\hat{b}_{\mu\nu}\tilde{B}^{\mu\nu}$ and $\hat{b}_{\mu\nu}=b_{\mu\nu}/\sqrt{b^{2}}$.
 
Hence, the equation of motion \eqref{eqmotion3} can be rewritten as
\ie\
\partial_{\mu}\tilde{G}^{\mu\nu\lambda} + \Box\beta\hat{b}^{\nu\lambda} + \partial_{\mu}\partial^{\lambda}\beta\hat{b}^{\mu\nu} + \partial_{\mu}\partial^{\nu}\beta\hat{b}^{\lambda\mu}+4\lambda b^{2}\beta\hat{b}^{\nu\lambda} = 0,\label{eqmotion4}
\fe where $\tilde{G}_{\mu\nu\lambda}\equiv \partial_{\mu}A_{\nu\lambda} + \partial_{\lambda} A_{\mu\nu} + \partial_{v} A_{\lambda\mu}$. After multiplying Eq. \eqref{eqmotion4} by $\partial_{\nu}$, we obtain the following constraint:
\ie
b_{\lambda\nu}\partial^{\nu}\beta = 0.\label{constrain1}
\fe 

Since the fields $A_{\mu\nu}$ and $\beta$ are independent, we can extract their respective equations of motion inserting the constraint \eqref{constrain1} into \eqref{eqmotion4} and applying the projections \eqref{projectors1}. Thus,
\begin{eqnarray}
&&\Box A_{\mu\nu} + \partial_{\mu}\partial^{\alpha}A_{\nu\alpha} + \partial_{\nu}\partial^{\alpha}A_{\alpha\mu} - \frac{2}{b^{2}} b_{\mu\nu} b_{\alpha\beta} \partial^{\beta}\partial_{\lambda}A^{\lambda\alpha} = 0, \label{EqA}\\
&&\Box \beta + 4\lambda b^{2} \beta + 2 \hat{b}_{\mu\nu}\partial^{\nu}\partial_{\alpha}A^{\alpha\mu} = 0.\label{Eqbeta}
\end{eqnarray}

These equations indicate that modes remain coupled and the dispersions relations, including the mass value, cannot be correctly identified yet. However, the transverse components of $A_{\mu\nu}$, namely, those satisfying the condition $\partial_{\mu}A^{\mu\nu} = 0$, remain unaffected even when the massive mode $\beta$ is nonzero. We can decouple the equations \eqref{EqA} and \eqref{Eqbeta} by noting that the constraint equation \eqref{constrain1} imposes for the massive mode $\beta$ the additional requirement
\ie
b_{\mu\nu}p^{\nu}=0,
\fe such that the energy-momentum vector associated with the massive mode is orthogonal to the vacuum value $b_{\mu\nu}$. This condition entails the following dispersion relation for the massive mode
\ie
p^{2}-4\lambda b^{2}=0,
\fe with the associated mass value given by $m_{\beta}^{2} \equiv 4\lambda b^{2}.$

At first glance, this relation might suggest that the massive mode has a physical mass when $4\lambda b^{2}$ is positive. However, it is possible to show that there is a special observer frame in which $b_{\mu\nu}$ assumes a simplified block-diagonal form
\begin{eqnarray}
b_{\mu\nu}=\begin{pmatrix}0 & -a & 0 & 0\\
a & 0 & 0 & 0\\
0 & 0 & 0 & d\\
0 & 0 & -d & 0
\end{pmatrix},\label{diag}
\end{eqnarray}such that $b^{2} = - 2 (a^{2}-d^{2}).$ Notice that the six parameters initially required to define $b_{\mu\nu}$ in an arbitrary referential are reduced to only two nonzero real numbers in this particular frame \cite{AntiSy-Altschul,Hernaski2016}. It is easy to see that this specific form for $b_{\mu\nu}$ combined with the constraint equation \eqref{constrain1} implies to $\partial_{\mu}\beta = 0$, and hence $\beta$ is a constant at linear order. To satisfy the asymptotic boundary conditions, this amplitude must be set to zero. We conclude then that there is no physical propagating massive mode in the spectrum of theory at leading order. The remaining $A_{\mu\nu}$ transverse mode propagates as a usual gauge Kalb-Ramond field, containing only one degree of freedom and behaving like a real scalar field \cite{AntiSy-Ogievetsky,Sezgin} . 

The above results are in agreement with those obtained in Ref. \cite{AntiSy-Altschul}, in which the authors explore the equivalence between the theory described by \eqref{LBV} and its dual correspondence defined by
\ie
\mathcal{L}_{A,B,V} = \frac{1}{2} B_{\mu\nu}\epsilon^{\mu\nu\alpha\beta}F_{\alpha\beta}+\frac{1}{2}A_{\mu}A^{\mu}-V, \label{LABV}
\fe where $F_{\mu\nu}$ is the field strength for a vector field $A_{\mu}$. It is worth mentioning that a similar analysis was carried out in the bumblebee electrodynamics in Ref. \cite{MalufRadiative}.

\section{Spontaneous Lorentz violation and the Kalb-Ramond propagator\label{KRpropagador}}

Using the decomposition \eqref{vacuum}, we can write the Lagrange density \eqref{LBV} as 
\ie
\mathcal{L}_{kin}=\frac{1}{6}\tilde{H}_{\mu\nu\alpha}\tilde{H}^{\mu\nu\alpha}
-2\lambda b_{\mu\nu}b_{\alpha\beta}\tilde{B}^{\mu\nu}\tilde{B}^{\alpha\beta},\label{LKR2}
\fe where we have collected only terms up to second order in $\tilde{B}_{\mu\nu}$. Clearly, the gauge symmetry was broken in \eqref{LKR2} due to terms which depend on the vacuum value $b_{\mu\nu}$.

To find the Feynman propagator and consequently the modifications ascribed to the Lorentz violation on the particle spectrum of the theory, we put the kinetic Lagrangian into the bilinear form
\ie
\mathcal{L}_{kin}=\frac{1}{2}\tilde{B}^{\mu\nu}\hat{\mathcal{O}}_{\mu\nu,\alpha\beta}\tilde{B}^{\alpha\beta},
\fe where the operator $\hat{\mathcal{O}}_{\mu\nu,\alpha\beta}$  is antisymmetric in the indices $(\mu\nu)$, $(\alpha\beta)$, and symmetric under interchange of pairs $(\mu\nu)$ and $(\alpha\beta)$. Then, this operator takes the form
\begin{eqnarray}
\hat{\mathcal{O}}_{\mu\nu,\alpha\beta}=-\frac{\Box}{2}(\eta_{\mu\alpha}\eta_{\nu\beta}-\eta_{\mu\beta}\eta_{\nu\alpha})-\frac{1}{2}(\partial_{\mu}\partial_{\beta}\eta_{\nu\alpha}-\partial_{\nu}\partial_{\beta}\eta_{\mu\alpha}-\partial_{\mu}\partial_{\alpha}\eta_{\nu\beta}+\partial_{\nu}\partial_{\alpha}\eta_{\mu\beta})-4\lambda b_{\mu\nu}b_{\alpha\beta}.\nonumber\\
\end{eqnarray}

Following this notation, the Feynman propagator is defined as 
\ie
\langle 0 |T\left[\tilde{B}_{\mu\nu}(x)\tilde{B}_{\alpha\beta}(y)\right]| 0 \rangle=i\left(\hat{\mathcal{O}}^{-1}\right)_{\mu\nu,\alpha\beta}\delta^{4}(x-y).
\fe 

To invert the operator $\hat{\mathcal{O}}$, it is convenient to expand this operator on a basis of tensor projectors which satisfy a closed algebra. As it is well known, the set of spin-projection operators for the Lorentz-invariant antisymmetric 2-tensors  are defined as \cite{Colato}
\begin{eqnarray}
P^{(1)}_{\mu\nu ,\alpha\beta}&=&\frac{1}{2}(\theta_{\mu\alpha}\theta_{\nu\beta}-\theta_{\mu\beta}\theta_{\nu\alpha}),\nonumber \\ 
P^{(2)}_{\mu\nu ,\alpha\beta}&=&\frac{1}{4}(\theta_{\mu\alpha}\omega_{\nu\beta}-\theta_{\nu\alpha}\omega_{\mu\beta}-\theta_{\mu\beta}\omega_{\nu\alpha}+\theta_{\nu\beta}\omega_{\mu\alpha}),\label{Projector12}
\end{eqnarray}where 
\ie 
\theta_{\mu\nu}=\eta_{\mu\nu}-\omega_{\mu\nu},\ \ \ \omega_{\mu\nu}=\frac{\partial_{\mu}\partial_{\nu}}{\Box},
\fe are the transverse and longitudinal operators for vectors, respectively.

The usual spin-projection operators satisfy the orthogonality relation
\ie
P^{(i)}_{\mu\nu ,\rho\sigma}P^{(j)}_{\rho\sigma,\alpha\beta}=\delta^{ij}P^{(i)}_{\mu\nu ,\alpha\beta},
\fe with $i,j=1,2$ and the tensorial completeness relation:
\ie
\left[P^{(1)}+P^{(2)}\right]_{\mu\nu ,\alpha\beta}=\frac{1}{2}(\eta_{\mu\alpha}\eta_{\nu\beta}-\eta_{\mu\beta}\eta_{\nu\alpha})=\mathcal{I}_{\mu\nu , \alpha\beta}.
\fe

One has to introduce additional operators to the two usual spin-projection operators defined in \eqref{Projector12} to accommodate the mass matrix generated by spontaneous Lorentz symmetry breaking. Our analysis of the spin operators generated by the Lorentz 
symmetry violation yields the new set of structures listed below:
\begin{eqnarray}
P^{(3)}_{\mu\nu ,\alpha\beta}&=&P^{\perp}_{\mu\nu , \alpha\beta},\\
P^{(4)}_{\mu\nu ,\alpha\beta}&=& \frac{1}{2}\left(  \omega_{\mu\lambda} \,P^{\parallel}_{\nu\lambda , \alpha\beta} - \omega_{\nu\lambda}\, P^{\parallel}_{\mu\lambda , \alpha\beta}    \right),\\
P^{(5)}_{\mu\nu ,\alpha\beta}&=&\frac{1}{2}\left(  \omega_{\alpha\lambda} \,P^{\parallel}_{\mu\nu , \beta\lambda} - \omega_{\beta\lambda}\, P^{\parallel}_{\mu\nu , \alpha\lambda}    \right),\\
P^{(6)}_{\mu\nu ,\alpha\beta}&=&\frac{1}{4}\left(  \omega_{\mu\alpha} \,P^{\parallel}_{\nu\rho , \beta\sigma}\,\omega^{\rho\sigma} - \omega_{\nu\alpha}\, P^{\parallel}_{\mu\rho , \beta\sigma}\,\omega ^{\rho\sigma} -\omega_{\mu\beta}\, P^{\parallel}_{\nu\rho , \alpha\sigma}\,\omega ^{\rho\sigma}  + \omega_{\nu\beta}\, P^{\parallel}_{\mu\rho , \alpha\sigma}\,\omega ^{\rho\sigma} \right),
\end{eqnarray}where $P^{\perp}$ and $P^{\parallel}$ were defined in \eqref{projectors1}.

These new operators together with the spin-projection operators \eqref{Projector12}
satisfy a closed algebra explicitly shown in Table \ref{table1} and Table \ref{table2}.
\begin{table}[ptb]
\begin{tabular}{|c|c|c|c|}
\hline 
 & $P^{(1)}$ & $P^{(2)}$ & $P^{(3)}$ \\ 
\hline 
$P^{(1)}$ & $P^{(1)}$ & $0$ & $P^{(3)}-P^{(2)}-2P^{(4)}$ \\ 
\hline 
$P^{(2)}$ & $0$ & $P^{(2)}$ & $P^{(2)}+2P^{(4)}$ \\ 
\hline 
$P^{(3)}$ & $P^{(3)}-P^{(2)}-2P^{(5)}$ & $P^{(2)}+2P^{(5)}$ & $P^{(3)}$ \\ 
\hline 
$P^{(4)}$ & $P^{(4)}+2P^{(6)}$ & $-2P^{(6)}$ & $0$ \\ 
\hline 
$P^{(5)}$ & $0$ & $P^{(5)}$ & $P^{(5)}+\frac{(b_{\mu\nu}p^{\nu})^{2}}{p^{2}b^{2}}(P^{(1)}+P^{(2)}-P^{(3)})$ \\ 
\hline 
$P^{(6)}$ & $0$ & $P^{(6)}$ & $P^{(6)}+\frac{(b_{\mu\nu}p^{\nu})^{2}}{p^{2}b^{2}}P^{(4)}$ \\ 
\hline 
\end{tabular} 
\caption{Algebra of tensor projectors.}%
\label{table1}%
\end{table}

\begin{table}[ptb]
\begin{tabular}{|c|c|c|c|}
\hline 
 & $P^{(4)}$ & $P^{(5)}$ & $P^{(6)}$ \\ 
\hline 
$P^{(1)}$ & $0$ & $P^{(5)}+2P^{(6)}$ & $0$ \\ 
\hline 
$P^{(2)}$ & $P^{(4)}$ & $-2P^{(6)}$ & $P^{(6)}$ \\ 
\hline 
$P^{(3)}$ & $P^{(4)}+\frac{(b_{\mu\nu}p^{\nu})^{2}}{p^{2}b^{2}}(P^{(1)}+P^{(2)}-P^{(3)})$ & $0$ & $P^{(6)}+\frac{(b_{\mu\nu}p^{\nu})^{2}}{p^{2}b^{2}}P^{(5)}$ \\ 
\hline 
$P^{(4)}$ & $-\frac{(b_{\mu\nu}p^{\nu})^{2}}{p^{2}b^{2}}P^{(4)}$ & $P^{(6)}$ & $-\frac{(b_{\mu\nu}p^{\nu})^{2}}{p^{2}b^{2}}P^{(6)}$ \\ 
\hline 
$P^{(5)}$ & $\frac{(b_{\mu\nu}p^{\nu})^{2}}{2p^{2}b^{2}}(P^{(1)}+P^{(2)}-P^{(3)})$ & $-\frac{(b_{\mu\nu}p^{\nu})^{2}}{p^{2}b^{2}}P^{(5)}$ & $-\frac{(b_{\mu\nu}p^{\nu})^{2}}{2p^{2}b^{2}}P^{(5)}$ \\ 
\hline 
$P^{(6)}$ & $\frac{(b_{\mu\nu}p^{\nu})^{2}}{2p^{2}b^{2}}P^{(4)}$ & $-\frac{(b_{\mu\nu}p^{\nu})^{2}}{p^{2}b^{2}}P^{(6)}$ & $\frac{(b_{\mu\nu}p^{\nu})^{2}}{2p^{2}b^{2}}P^{(6)}$ \\ 
\hline 
\end{tabular} 
\caption{Algebra of tensor projectors.}%
\label{table2}%
\end{table}

Now we are ready to calculate the propagator. Let us write both operators 
$\hat{\mathcal{O}}$ and $\hat{\mathcal{O}}^{-1}$ as a linear combination of the  projectors $\{ P^{(1)},P^{(2)},P^{(3)},P^{(4)},P^{(5)},P^{(6)} \}$, such that
\begin{eqnarray}
\hat{\mathcal{O}}&=&x_{1}P^{(1)}+x_{2}P^{(2)}+x_{3}P^{(3)}+x_{4}P^{(4)}+x_{5}P^{(5)}+x_{6}P^{(6)},\nonumber\\
\hat{\mathcal{O}}^{-1}&=&y_{1}P^{(1)}+y_{2}P^{(2)}+y_{3}P^{(3)}+y_{4}P^{(4)}+y_{5}P^{(5)}+y_{6}P^{(6)},
\end{eqnarray}with the coefficients $x_{i}$ being scalar functions from the momentum and the $vev$ $b_{\mu\nu}$, and $y_{i}$ are coefficients to be determined.

For our specific case, the operator $\hat{\mathcal{O}}$ can be expanded in the form
\ie
\hat{\mathcal{O}}_{\mu\nu,\alpha\beta}=(-\Box-4\lambda b^{2}) P^{(1)}_{\mu\nu ,\alpha\beta}-4\lambda b^{2} P^{(2)}_{\mu\nu ,\alpha\beta}+4\lambda b^{2} P^{(3)}_{\mu\nu ,\alpha\beta}.
\fe
Taking into account that $\hat{\mathcal{O}}\hat{\mathcal{O}}^{-1}=\mathcal{I}$, and after performing the necessary algebra which is shown in Table \ref{table1} and Table \ref{table2}, we find the following result in the momentum space,
\ie
\hat{\mathcal{O}}^{-1}_{\mu\nu,\alpha\beta}=\frac{1}{p^{2}} P^{(1)}_{\mu\nu , \alpha\beta} + \frac{b^{2}}{(b_{\rho\sigma}p^{\sigma})^{2}}( P^{(4)}_{\mu\nu , \alpha\beta}+ P^{(5)}_{\mu\nu , \alpha\beta})\label{propagator2},
\fe where only the finite and non-zero coefficients were displayed.

The poles of the propagator determine the particle content of the
model. In expression \eqref{propagator2}, we noticed the presence of
two distinct poles:
\begin{eqnarray}
p^{2} & = & 0,\\
(b_{\mu\nu}p^{\nu})^{2} & = & 0.
\end{eqnarray}

The first pole $p^{2}=0$ confirms the presence of a massless excitation,
which can be identified as one NG mode of the $A_{\mu\nu}$. The
breaking of Lorentz symmetry lies in the pole $(b_{\mu\nu}p^{\nu})^{2}=0$.
For a better comprehension of the structure of this pole, we may use the background
tensor $b_{\mu\nu}$ in the simple block-diagonal form \eqref{diag}. From this
choice, we can write $(b_{\mu\nu}p^{\nu})^{2}=0$ as
\begin{equation}
\left(p^{0}\right)^{2}-\left(p^{1}\right)^{2}+\frac{d^{2}}{a^{2}}\left(\left(p^{2}\right)^{2}+\left(p^{3}\right)^{2}\right)=0,
\end{equation}
which indeed represents a massless mode propagating in an anisotropic
medium, and can be identified with the others NG modes of the $A_{\mu\nu}$.
From this relation, we can analyze the energy stability and causality
of this mode. The dispersion relation is
\begin{equation}
p^{0}=\pm\sqrt{\left(p^{1}\right)^{2}-\frac{d^{2}}{a^{2}}\left(\left(p^{2}\right)^{2}+\left(p^{3}\right)^{2}\right)}.\label{dispersion}
\end{equation}

The minus sign of Eq. \eqref{dispersion} indicates an instability of the energy spectrum
of the model. In the case of $b^{2}=2(d^{2}-a^{2})>0$, the theory exhibits
tachyons in the particle spectrum. At the classical level, this non-physical
pole does not modify the interparticle potential, since the Lorentz-symmetry
violating terms associated with the projection operators $P^{(4)}$
and $P^{(5)}$ does not contribute to any observable associated with
the $S$-matrix at tree-level approximation (due to the conservation
of external currents). However, quantum effects could excite such mode, and a convenient prescription in the momentum integrals
should be implemented to deal with this pole. These issues considering stability under radiative corrections lie beyond the scope of the present work.

Finally, we can conclude that only massless poles associated with
the NG modes, contained in the transverse component $A_{\mu\nu}$, were
generated by the spontaneous Lorentz symmetry breaking, and therefore, the
massive mode $\beta$ is non-propagating at leading order. These results
are in agreement with those obtained above and in Ref. \cite{AntiSy-Altschul}.

\section{Conclusion\label{conclusion}}

We have considered the spontaneous Lorentz symmetry breaking due to an anti-symmetric 2-tensor field (or Kalb-Ramond field) triggered by a smooth quadratic potential in Minkowski spacetime. The resulting solutions to the equations of motion for the free field contain massless NG and massive modes which appear mixed by a mass matrix. We have shown that these solutions obey some constraints that lead to the massive mode be non-propagating at leading order. Such results are in agreement with the previous work \cite{AntiSy-Altschul}. Furthermore, we have evaluated the modified Kalb-Ramond propagator by the presence of Lorentz-violating terms using a new algebra of spin-projection operators. The analysis of the propagator poles revealed that no physical mass was generated by the spontaneous Lorentz symmetry breaking and that the non-physical modes could not modify the interparticle potential at leading order. The determination of the exact form of the Kalb-Ramond propagator allows the application of the tensor calculation techniques for some interesting problems. The issue whether the massive mode propagates at higher orders is an interesting open question \cite{AntiSy-Altschul}, and the calculation of the radiative corrections can help to elucidate this subject. Moreover, we may use the $\tilde{B}_{\mu\nu}$ propagator to access corrections at higher orders in the gravitational scenario \cite{MalufGravity1,MalufGravity2}. Some investigations in this direction are now under development.


\section*{Acknowledgments}
\hspace{0.5cm}The authors would like to thank the Funda\c{c}\~{a}o Cearense de apoio ao Desenvolvimento
Cient\'{\i}fico e Tecnol\'{o}gico (FUNCAP), the Coordena\c{c}\~ao de Aperfei\c{c}oamento de Pessoal de N\'ivel Superior (CAPES), and the Conselho Nacional de Desenvolvimento Cient\'{\i}fico e Tecnol\'{o}gico (CNPq) for
financial support.

\end{document}